\documentclass[11pt]{article}

\usepackage[margin=1in]{geometry}
\usepackage[T1]{fontenc}
\usepackage[utf8]{inputenc}
\usepackage{lmodern}
\usepackage{microtype}
\usepackage{setspace}

\usepackage{authblk}

\usepackage{amsmath,amssymb}
\usepackage{siunitx}
\usepackage{graphicx}
\usepackage{booktabs}
\usepackage{array}
\usepackage{tabularx}
\usepackage{adjustbox}
\usepackage{multirow}
\usepackage{float}
\usepackage{placeins}
\usepackage{caption}
\usepackage{subcaption}

\usepackage{enumitem}
\setlist[itemize]{leftmargin=1.5em}
\setlist[enumerate]{leftmargin=1.5em}

\usepackage{xcolor}
\usepackage[numbers,sort&compress]{natbib}
\usepackage{hyperref}
\usepackage[nameinlink,noabbrev]{cleveref}

\hypersetup{
    colorlinks=true,
    linkcolor=blue!50!black,
    citecolor=blue!50!black,
    urlcolor=blue!50!black,
    pdftitle={A Reproducible AAVSO Johnson-V Fourier Template for the Prototype Cepheid Delta Cephei},
    pdfauthor={Zuhoor Elahi, Christopher Sirola, Wafa Gull}
}

\graphicspath{{./}{figures/}}

\newcommand{\dcep}{Delta Cephei}
\newcommand{\Pobs}{P_{\rm obs}}
\newcommand{\Vband}{Johnson-$V$}
\newcommand{\Deltav}{\Delta V}
\newcommand{\frise}{f_{\rm rise}}
\newcommand{\Aasym}{A_{\rm asym}}
\newcommand{\Rtwone}{R_{21}}
\newcommand{\Rthreeone}{R_{31}}
\newcommand{\phitwone}{\phi_{21}}
\newcommand{\phithreeone}{\phi_{31}}

\title{A Reproducible AAVSO Johnson-$V$ Fourier Template for the Prototype Cepheid Delta Cephei}

\author[1,2,*]{Zuhoor Elahi}
\author[1]{Wafa Gull}

\affil[1]{Department of Physics and Astronomy, University of Southern Mississippi, Hattiesburg, MS, USA}
\affil[2]{Department of Physics, University of Karachi, Karachi, Pakistan}
\affil[*]{Corresponding author: zuhoor.elahi@usm.edu}

\date{}

\begin{document}

\maketitle

\begin{abstract}
We present an empirical Fourier reconstruction of the observed \Vband\ light curve of the prototype Classical Cepheid \dcep. The goal is not to infer a full physical stellar model, but to establish a reproducible observed-light-curve benchmark for later comparison with nonlinear pulsation, synthetic-photometry, Baade--Wesselink/SPIPS, GYRE-supported, and independent hydrodynamic calculations. Using an adopted period of $\Pobs=5.366531\,{\rm d}$, 244 AAVSO \Vband\ measurements were filtered to a cleaned sample of 242 points after rejecting two extreme outliers. The cleaned data span 355.09259 d and were phase folded using an empirical bright-maximum epoch of ${\rm JD}=2460851.395800$. We fit a low-order Fourier model to the phased light curve and adopt a third-order template as the preferred empirical morphology representation. The adopted $N=3$ fit gives $A_0=3.9031^{+0.0064}_{-0.0061}$, $A_1=0.3434^{+0.0083}_{-0.0093}$ mag, $A_2=0.1428^{+0.0089}_{-0.0084}$ mag, and $A_3=0.0531^{+0.0080}_{-0.0075}$ mag, corresponding to $\Rtwone=0.4159^{+0.0303}_{-0.0262}$ and $\Rthreeone=0.1547^{+0.0237}_{-0.0227}$. The template has a full \Vband\ amplitude of $\Deltav=0.8544^{+0.0225}_{-0.0222}$ mag, a rise fraction of $\frise=0.2885^{+0.0080}_{-0.0070}$, and an asymmetry index of $\Aasym=0.4230^{+0.0140}_{-0.0160}$. Fourier orders $N=4$--6 reduce the RMS residual by only approximately $0.0012$ mag relative to the $N=3$ model, so the third-order representation is retained as the simplest adequate empirical template. Observer-level residual diagnostics show that the remaining scatter is not purely random: the four principal observer groups have mean residuals ranging from $-0.0870$ to $+0.1025$ mag. This paper provides an observational morphology target rather than a physical explanation of the pulsation.
\end{abstract}


\section{Introduction}

Classical Cepheids are central to stellar pulsation studies and to the calibration of the extragalactic distance scale. The period--luminosity relation was established from the discovery that longer-period Cepheid-like variables in the Small Magellanic Cloud are systematically brighter \citep{leavitt1912}. This empirical relation has since been developed into a major distance-scale tool through Galactic calibrations, extragalactic Cepheid surveys, and modern parallax-based work \citep{madorefreedman1991,freedman2001,benedict2007,riess2016,riess2022}. As the prototype of the class, \dcep\ remains an important reference object for testing empirical and theoretical descriptions of Cepheid variability.

The present work focuses only on the observed optical light-curve morphology of \dcep. As a nearby prototype Cepheid, \dcep\ has been studied both as an individual pulsating star and as a calibrator for Cepheid-based distance work \citep{benedict2002,benedict2007,merand2005}. The American Association of Variable Star Observers (AAVSO) describes \dcep\ as the prototype Cepheid variable, with a visual range of approximately 3.5--4.4 mag and a period near 5.366 d \citep{aavso_delta_cep}. These values make the star well suited for a direct empirical reconstruction of its phase-dependent brightness variation. In contrast to physical modeling papers that attempt to compute the pulsation from stellar structure, convection, opacity, or hydrodynamics, the present paper treats the observed Johnson-$V$ (visual-band) light curve itself as the primary object of study.

Fourier decomposition provides a compact and reproducible way to describe Cepheid light-curve shapes. The use of harmonic amplitudes, amplitude ratios, and phase combinations to quantify Cepheid morphology has a long history in optical light-curve analysis \citep{simonlee1981,simonmoffett1985,antonello1986,andreasen1988}. Later studies extended Fourier and template-based analysis to larger samples, multiple photometric bands, and automated or semi-automated light-curve characterization \citep{kanburngeow2002,ngeow2003,ngeowkanbur2004,tanvir2005,debsingh2009,bhardwaj2015,bhardwaj2017,soszynski2008}. These quantities are useful because they separate several aspects of the observed morphology: the total amplitude, the strength of the first few harmonics, the relative phase of the harmonics, the rapidity of the rise to maximum light, and the asymmetry between the rising and declining branches.

The novelty of the present contribution is therefore not the Fourier method itself. Instead, the contribution is a reproducible AAVSO-based \Vband\ benchmark for the prototype Cepheid using explicit data cleaning, harmonic-order testing, bootstrap uncertainties, and observer-level residual diagnostics. This framing is important because the literature already contains extensive Cepheid Fourier analyses, but a compact and reproducible Johnson-$V$ benchmark tied to the present AAVSO dataset is useful for direct comparison with model light curves in future empirical and theoretical studies.

The motivation for this paper is therefore simple. Before comparing physical models of \dcep\ with the observations, it is useful to define a clean empirical target. A nonlinear pulsation model, synthetic light curve, Spectro-Photo-Interferometry of Pulsating Stars (SPIPS)/Baade--Wesselink reconstruction, or other hydrodynamic calculation may reproduce the observed period while still failing to reproduce the optical amplitude or light-curve morphology. This paper provides the observed \Vband\ morphology benchmark against which such models may be compared.

This study does not attempt to determine the mass, luminosity, evolutionary crossing, opacity table, convection parameters, or radius variation of \dcep. It also does not perform an $O-C$ period-change analysis. Those are physical or evolutionary questions. Here we construct a phase-folded \Vband\ template, quantify its Fourier morphology, estimate its uncertainties, examine residual structure, and test whether higher Fourier orders materially improve the empirical representation.

\section{Relation to Previous Cepheid Fourier Work}

The present analysis follows a well-established methodological tradition. Early Fourier studies showed that Cepheid light curves can be described by harmonic amplitudes and phase combinations and that these quantities carry information about the structure and morphology of the pulsation cycle \citep{simonlee1981,simonmoffett1985}. Subsequent work refined the use of Fourier parameters for Cepheid classification, light-curve comparison, and multi-band morphology studies \citep{antonello1986,andreasen1988,ngeow2003,debsingh2009,bhardwaj2015}. Large photometric surveys, including OGLE, have made Fourier descriptors especially useful because they provide compact summaries of thousands of variable-star light curves \citep{soszynski2008}.

Table~\ref{tab:literature_context} summarizes the role of several representative studies relative to the present paper. The purpose is not to provide a complete review of Cepheid Fourier analysis, but to clarify the scope of this work. The present paper is narrower than large-sample Fourier surveys and is not intended to introduce a new mathematical method. Its purpose is to provide a transparent, reproducible, single-star \Vband\ benchmark for \dcep\ that can be used directly in later model comparisons.

\begin{table}[htbp]
\centering
\caption{Representative literature context for the present empirical Fourier-template analysis.}
\label{tab:literature_context}
\begin{tabularx}{\linewidth}{p{3.0cm}p{3.0cm}X}
\toprule
Study & Main emphasis & Relevance to this paper \\
\midrule
\citet{simonlee1981} & Fourier structure of Cepheid light curves & Established harmonic decomposition as a morphology diagnostic for Cepheid light curves. \\
\citet{simonmoffett1985} & Classical Cepheid light-curve morphology & Provided further Fourier-based analysis of classical Cepheid light curves. \\
\citet{antonello1986,andreasen1988} & Cepheid Fourier parameters & Demonstrated the diagnostic use of precise Fourier parameters for Cepheid morphology. \\
\citet{ngeow2003,ngeowkanbur2004,tanvir2005} & Template and Fourier light-curve methods & Developed and applied Fourier/template methods to Cepheid light-curve reconstruction and parameterization. \\
\citet{debsingh2009} & Fourier and principal-component analysis & Showed the use of Fourier quantities in broader variable-star light-curve analysis. \\
\citet{bhardwaj2015,bhardwaj2017} & Multi-band observed and theoretical Cepheid morphology & Placed Fourier parameters in a multiwavelength observational and model-comparison context. \\
\citet{soszynski2008} & Large survey catalog of Cepheids & Illustrates the role of homogeneous survey light curves and Fourier-like descriptors in modern variable-star studies. \\
This work & AAVSO Johnson-$V$ template for \dcep & Provides a reproducible single-star benchmark with data cleaning, order testing, bootstrap uncertainties, and observer-level residual diagnostics. \\
\bottomrule
\end{tabularx}
\end{table}

\section{Observational Data}

The input data are AAVSO \Vband\ observations of \dcep\ obtained from the AAVSO International Database (AID) and cited following AAVSO data-use guidance \citep{aavso_aid,aavso_data_guidelines}. AAVSO observations of \dcep\ have also been discussed in the context of light-curve quality, visual observing practice, and period-change work \citep{turner2011}. The raw file contains 244 measurements. All points are in the \Vband\ bandpass. Two measurements with magnitudes near $V\sim6.3$ were rejected as extreme outliers relative to the observed Cepheid light curve. The final cleaned sample contains 242 observations from six observers. The cleaned data span 355.09259 d, from Julian Date (JD) 2460821.784320 to JD 2461176.876910.

Table~\ref{tab:data_summary} summarizes the observational sample used in the analysis. The raw cleaned magnitude range is $V=3.353$ to $V=4.443$ mag. This raw point-to-point range is larger than the amplitude of the fitted Fourier template because it includes measurement scatter, observer-to-observer offsets, and individual points not lying exactly on the mean phased curve.

\begin{table}[htbp]
\centering
\caption{Summary of the AAVSO \Vband\ data used in this work.}
\label{tab:data_summary}
\begin{tabular}{lc}
\toprule
Quantity & Value \\
\midrule
Raw measurements & 244 \\
Johnson-$V$ measurements & 244 \\
Cleaned measurements & 242 \\
Number of observers & 6 \\
JD minimum & 2460821.784320 \\
JD maximum & 2461176.876910 \\
Time span & 355.09259 d \\
Time span & 0.97219 yr \\
Brightest cleaned point & $V=3.353$ mag \\
Faintest cleaned point & $V=4.443$ mag \\
Raw cleaned magnitude range & 1.090 mag \\
Adopted period & $5.366531$ d \\
Empirical bright-maximum epoch & JD 2460851.395800 \\
\bottomrule
\end{tabular}
\end{table}

The AAVSO data are heterogeneous in the sense that they are obtained by multiple observers and may include different instrumental and transformation histories. This heterogeneity is not treated as a defect in the present paper; it is part of the reason the residual analysis is included. The purpose is to construct a robust empirical morphology template, not a sub-millimagnitude precision photometric standard.

\section{Phase Folding and Fourier Reconstruction}

The cleaned observations were folded using the adopted pulsation period

\begin{equation}
\Pobs = 5.366531\,{\rm d}.
\end{equation}

The pulsation phase was computed as

\begin{equation}
\phi = \left[\frac{t-T_0}{\Pobs}\right]\bmod 1,
\end{equation}

where $t$ is the observation time and $T_0$ is an empirical bright-maximum epoch. For the present dataset we use

\begin{equation}
T_0 = 2460851.395800.
\end{equation}

\begin{figure}[htbp]
\centering
\includegraphics[width=0.95\linewidth]{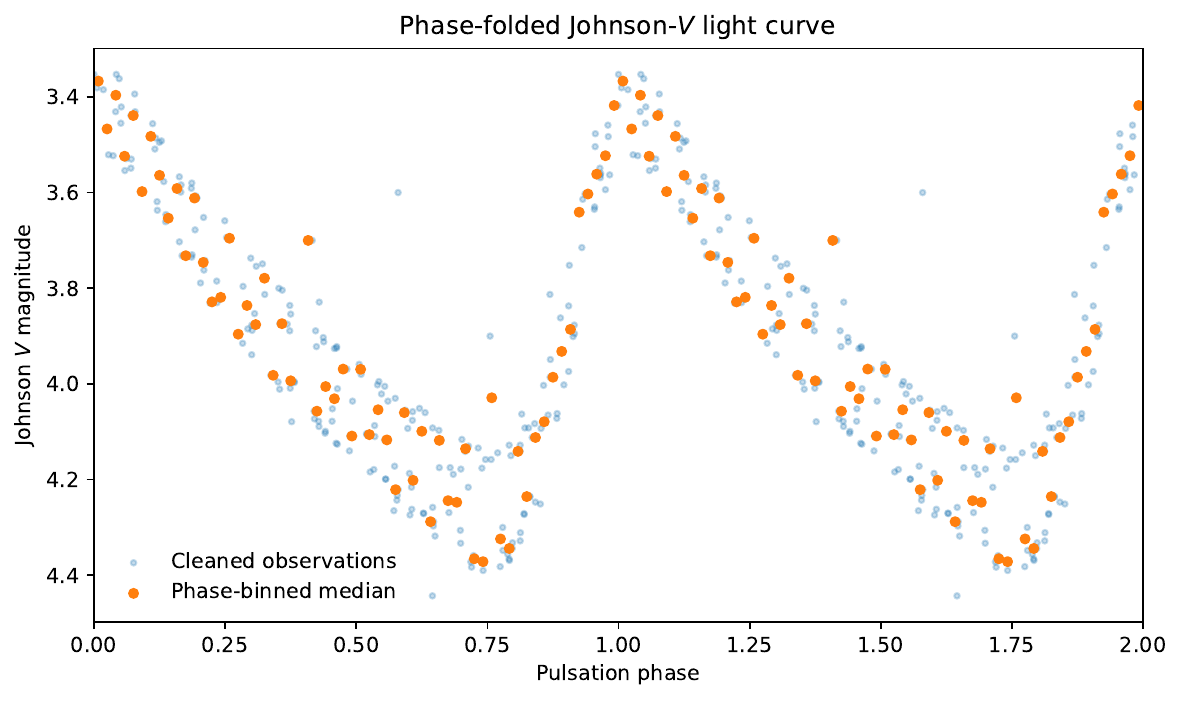}
\caption{Phase-folded AAVSO \Vband\ light curve of \dcep\ using $\Pobs=5.366531\,{\rm d}$. The data are repeated over two cycles for visual continuity. Blue points show the cleaned observations, and orange points show phase-binned medians. The magnitude axis is inverted so that brighter phases appear higher in the figure.}
\label{fig:phased_light_curve}
\end{figure}

The phased light curve is shown in Figure~\ref{fig:phased_light_curve}. The data are plotted over two cycles for visual continuity. The orange points show phase-binned medians, which trace the mean Cepheid morphology and reduce the visual effect of individual outliers.

The empirical template was fit with a Fourier series of the form

\begin{equation}
V(\phi) = A_0 + \sum_{k=1}^{N}
A_k\cos\left(2\pi k\phi+\varphi_k\right),
\label{eq:fourier_amp_phase}
\end{equation}

where $A_0$ is the mean magnitude, $A_k$ is the amplitude of harmonic order $k$, and $\varphi_k$ is the corresponding phase. In practice, the fit was performed as a linear least-squares problem using the equivalent form

\begin{equation}
V(\phi)=c_0+\sum_{k=1}^{N}
\left[c_k\cos(2\pi k\phi)+s_k\sin(2\pi k\phi)\right].
\label{eq:fourier_linear}
\end{equation}

The fitted sine and cosine coefficients were converted to amplitude and phase through

\begin{equation}
A_k = \sqrt{c_k^2+s_k^2},
\end{equation}

and

\begin{equation}
\varphi_k = \tan^{-1}\left(\frac{-s_k}{c_k}\right),
\end{equation}

using the two-argument arctangent to preserve the correct quadrant.

The principal Fourier amplitude ratios are

\begin{equation}
R_{k1} = \frac{A_k}{A_1}.
\end{equation}

In particular,

\begin{equation}
\Rtwone = \frac{A_2}{A_1},
\qquad
\Rthreeone = \frac{A_3}{A_1}.
\end{equation}

The Fourier phase combinations are defined as

\begin{equation}
\phi_{k1} = \varphi_k-k\varphi_1,
\end{equation}

wrapped to the interval $[0,2\pi)$. The reported phase combinations are $\phitwone$ and $\phithreeone$.

\section{Adopted Empirical Fourier Template}

\begin{figure}[htbp]
\centering
\includegraphics[width=0.95\linewidth]{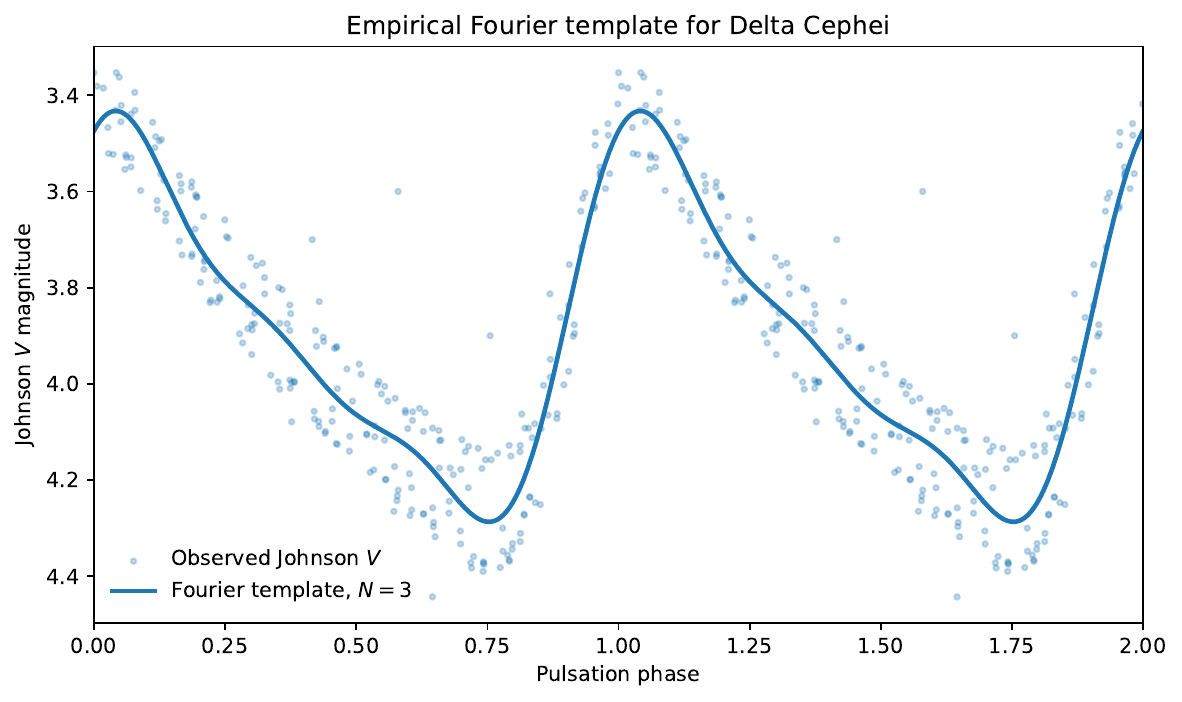}
\caption{Adopted empirical Fourier template for \dcep. The blue points are the cleaned AAVSO \Vband\ observations and the solid curve is the adopted third-order Fourier template. The fit is intended as a low-order morphology template rather than a precision model for every individual observation.}
\label{fig:template}
\end{figure}

Figure~\ref{fig:template} shows the adopted $N=3$ Fourier template overlaid on the observed phase-folded light curve. The template captures the dominant morphology: a relatively rapid rise to maximum brightness and a slower decline toward minimum brightness. The fit is not intended to represent every individual data point. Rather, it provides a smooth empirical description of the mean \Vband\ light curve.

The adopted Fourier coefficients are listed in Table~\ref{tab:fourier_coefficients}. The first harmonic has amplitude $A_1=0.343421$ mag. The second and third harmonics have amplitudes $A_2=0.142820$ mag and $A_3=0.053128$ mag, corresponding to $\Rtwone=0.415874$ and $\Rthreeone=0.154702$.

\begin{table}[htbp]
\centering
\caption{Adopted $N=3$ Fourier coefficients for the empirical \Vband\ template.}
\label{tab:fourier_coefficients}
\begin{tabular}{cccccc}
\toprule
$k$ & $c_k$ & $s_k$ & $A_k$ & $\varphi_k$ (rad) & $R_{k1}$ \\
\midrule
1 & $-0.242286$ & $-0.243384$ & $0.343421$ & $2.353934$ & $1.000000$ \\
2 & $-0.133974$ & $-0.049481$ & $0.142820$ & $2.787800$ & $0.415874$ \\
3 & $-0.052745$ & $0.006366$ & $0.053128$ & $-3.021483$ & $0.154702$ \\
\bottomrule
\end{tabular}
\end{table}

The mean magnitude coefficient is

\begin{equation}
A_0 = 3.903085.
\end{equation}

The resulting Fourier phase combinations are

\begin{equation}
\phitwone = 4.363117\,{\rm rad} = 249.988^\circ,
\end{equation}

and

\begin{equation}
\phithreeone = 2.483084\,{\rm rad} = 142.270^\circ.
\end{equation}

\begin{figure}[htbp]
\centering
\includegraphics[width=0.80\linewidth]{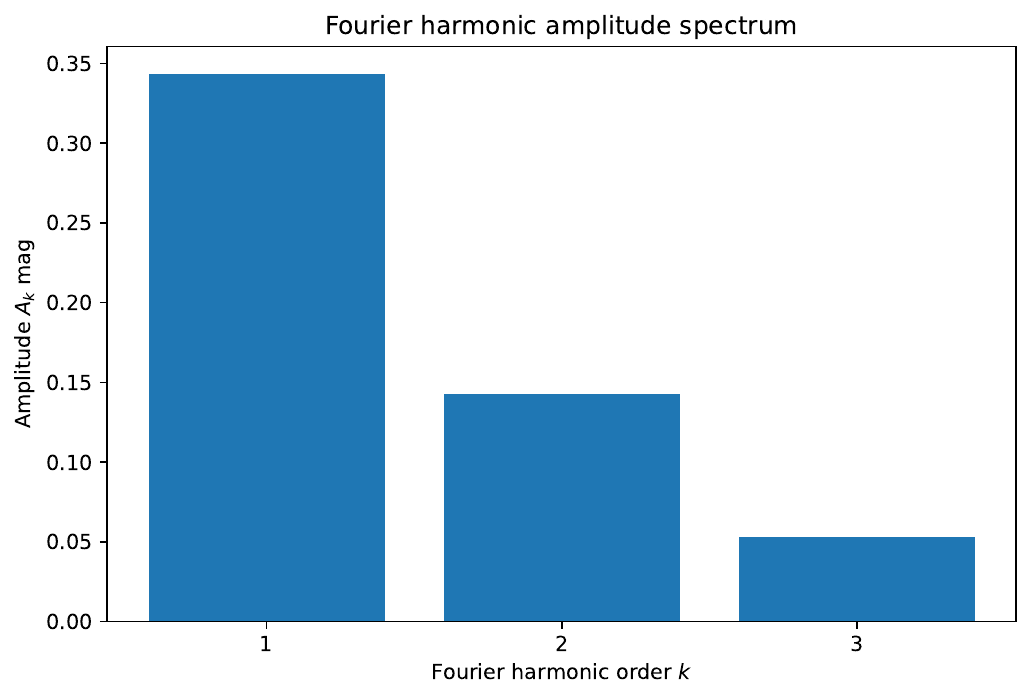}
\caption{Fourier harmonic amplitude spectrum for the adopted $N=3$ empirical template. The nonzero $A_2$ and $A_3$ terms quantify the departure of the observed light curve from a pure sinusoid.}
\label{fig:amplitude_spectrum}
\end{figure}

Figure~\ref{fig:amplitude_spectrum} shows the harmonic amplitude spectrum. The decreasing amplitudes with harmonic order indicate that the first harmonic dominates the total variation, while the second and third harmonics encode the non-sinusoidal morphology.

\section{Morphology Diagnostics}

The full template amplitude is defined as

\begin{equation}
\Deltav = V_{\rm faint}-V_{\rm bright},
\end{equation}

where $V_{\rm bright}$ is the minimum value of the magnitude curve and $V_{\rm faint}$ is the maximum value of the magnitude curve. For the adopted $N=3$ template,

\begin{equation}
\Deltav = 0.854439\,{\rm mag}.
\end{equation}

We define the rise fraction $\frise$ as the phase interval from faint minimum to bright maximum, normalized by the full pulsation cycle:

\begin{equation}
\frise = (\phi_{\rm bright}-\phi_{\rm faint}) \bmod 1.
\end{equation}

The decline fraction is then

\begin{equation}
f_{\rm decline}=1-\frise.
\end{equation}

The asymmetry index is defined as

\begin{equation}
\Aasym =
\frac{f_{\rm decline}-\frise}
{f_{\rm decline}+\frise}.
\end{equation}

Since $f_{\rm decline}+\frise=1$, this is equivalent to

\begin{equation}
\Aasym = 1-2\frise.
\end{equation}

For the adopted template,

\begin{equation}
\frise=0.2885,
\end{equation}

and

\begin{equation}
\Aasym=0.4230.
\end{equation}

These values quantify the classical Cepheid morphology: a relatively rapid rise to maximum brightness followed by a slower decline.

\begin{figure}[htbp]
\centering
\includegraphics[width=0.95\linewidth]{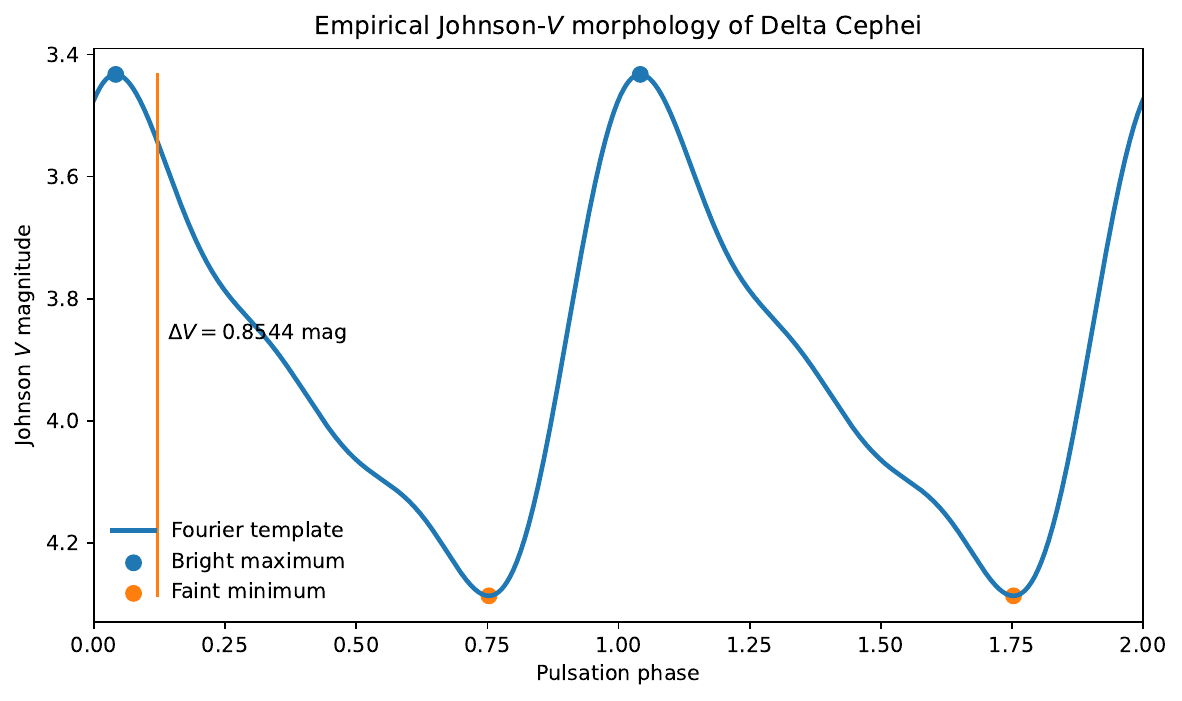}
\caption{Morphology summary for the adopted empirical \Vband\ Fourier template. The bright maximum corresponds to the minimum value of the magnitude curve, while the faint minimum corresponds to the maximum value. The adopted $N=3$ template gives $\Deltav=0.854439$ mag, $\frise=0.2885$, and $\Aasym=0.4230$.}
\label{fig:morphology_summary}
\end{figure}

Table~\ref{tab:morphology} summarizes the main morphology diagnostics. These quantities are the primary empirical benchmark produced by the present paper.

\begin{table}[htbp]
\centering
\caption{Main empirical morphology diagnostics for the adopted $N=3$ \Vband\ template.}
\label{tab:morphology}
\begin{tabular}{lc}
\toprule
Diagnostic & Value \\
\midrule
Adopted Fourier order & $N=3$ \\
$A_0$ & 3.903085 \\
$A_1$ & 0.343421 mag \\
$A_2$ & 0.142820 mag \\
$A_3$ & 0.053128 mag \\
$\Rtwone$ & 0.415874 \\
$\Rthreeone$ & 0.154702 \\
$\phitwone$ & 4.363117 rad = 249.988$^\circ$ \\
$\phithreeone$ & 2.483084 rad = 142.270$^\circ$ \\
$\Deltav$ & 0.854439 mag \\
$\frise$ & 0.2885 \\
$\Aasym$ & 0.4230 \\
Root-mean-square (RMS) residual & 0.096996 mag \\
Median absolute deviation (MAD) residual & 0.080869 mag \\
\bottomrule
\end{tabular}
\end{table}

\section{Residual Structure}

The residuals were computed as

\begin{equation}
\Delta V_{\rm res}(\phi)=V_{\rm obs}(\phi)-V_{\rm Fourier}(\phi).
\end{equation}

\begin{figure}[htbp]
\centering
\includegraphics[width=0.95\linewidth]{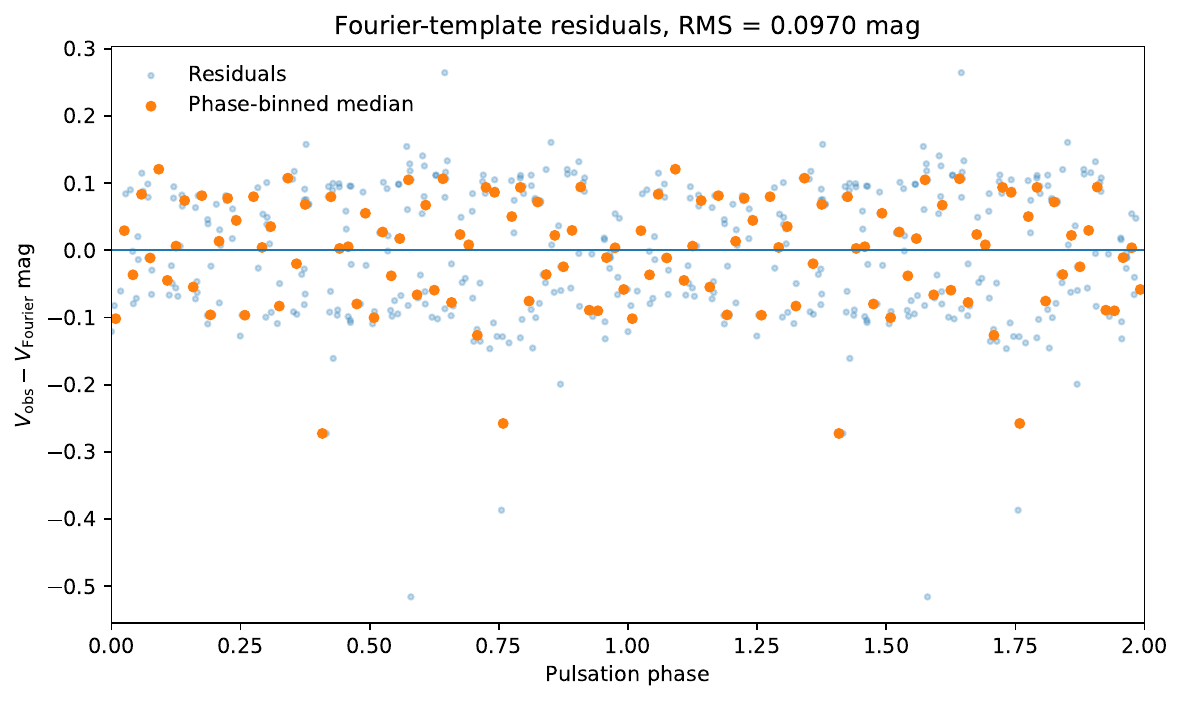}
\caption{Residuals of the adopted $N=3$ Fourier template. The residuals are defined as $V_{\rm obs}-V_{\rm Fourier}$. The RMS residual is 0.096996 mag. The phase-binned medians show that some phase-dependent structure remains, as expected for a low-order fit to heterogeneous observations.}
\label{fig:residuals}
\end{figure}

Figure~\ref{fig:residuals} shows the residuals as a function of phase. The RMS residual is 0.096996 mag, while the median absolute deviation is 0.080869 mag. The median residual is 0.006955 mag, indicating no large global offset. The maximum absolute residual is 0.515884 mag.

The residual scatter should not be interpreted as a failure of the empirical approach. The input dataset combines observations from multiple observers over nearly one year. The residuals may include measurement uncertainties, zero-point differences, transformation differences, atmospheric effects, and the fact that the adopted $N=3$ model intentionally avoids overfitting. Therefore, the template is best described as a low-order empirical morphology representation rather than a precision photometric standard.

\section{Fourier-Order Comparison}

To test whether the adopted third-order model is adequate, we repeated the analysis for Fourier orders $N=4$, $N=5$, and $N=6$. The results are summarized in Table~\ref{tab:order_comparison}. Higher orders reduce the RMS residual only slightly. The improvement from $N=3$ to $N=6$ is

\begin{equation}
0.096996-0.095793 = 0.001203\,{\rm mag}.
\end{equation}

This is approximately 1.2 mmag, much smaller than the observed residual scatter. The median absolute deviation is actually smallest for the $N=3$ fit. Therefore, there is no strong empirical justification for adopting a higher-order Fourier model as the main template.

\begin{table}[htbp]
\centering
\caption{Comparison of Fourier orders. The small change in RMS residual from $N=3$ to $N=6$ supports the use of the simpler third-order template as the adopted empirical morphology model.}
\label{tab:order_comparison}
\begin{tabular}{cccccc}
\toprule
$N$ & $\Deltav$ & $\frise$ & $\Aasym$ & RMS & MAD \\
 & mag & & & mag & mag \\
\midrule
3 & 0.854439 & 0.2885 & 0.4230 & 0.096996 & 0.080869 \\
4 & 0.842826 & 0.2690 & 0.4620 & 0.096183 & 0.084647 \\
5 & 0.825733 & 0.2620 & 0.4760 & 0.095829 & 0.084096 \\
6 & 0.829640 & 0.2640 & 0.4720 & 0.095793 & 0.083580 \\
\bottomrule
\end{tabular}
\end{table}

\begin{figure}[htbp]
\centering
\includegraphics[width=0.95\linewidth]{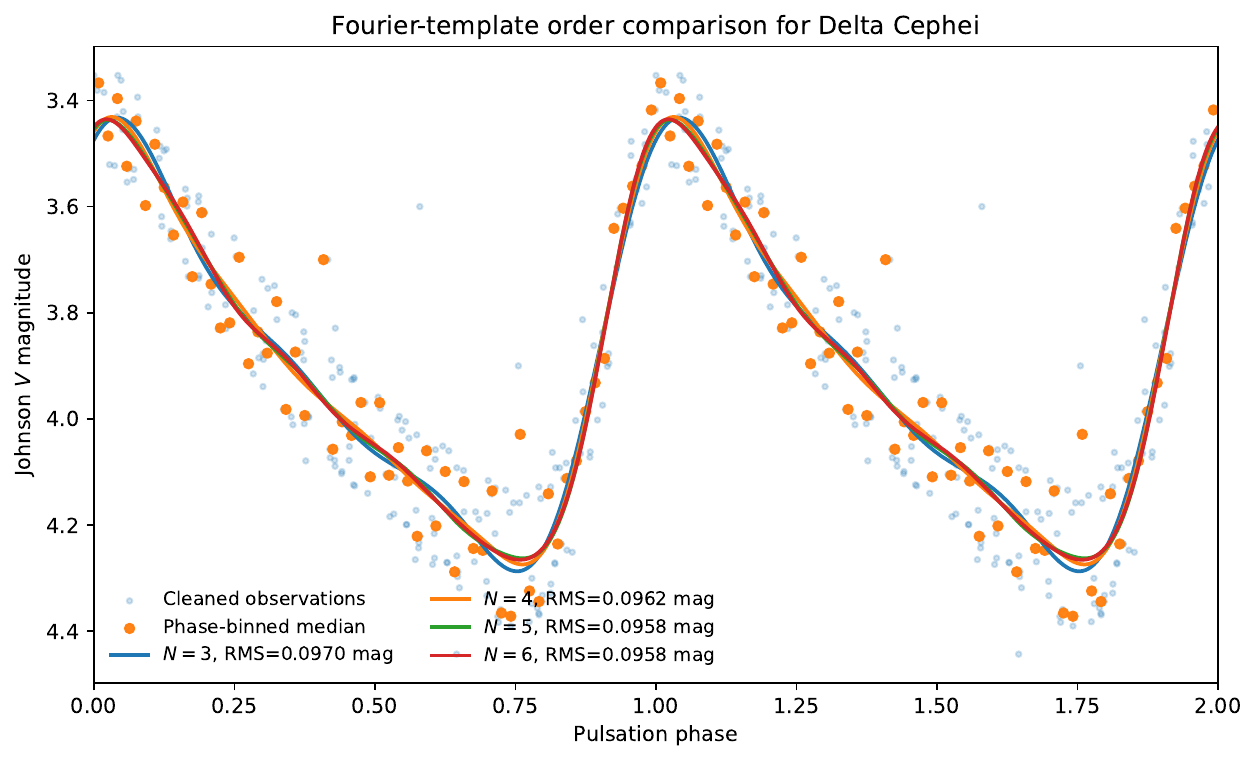}
\caption{Comparison of Fourier templates with $N=3$, 4, 5, and 6. The higher-order fits reduce the RMS residual only slightly and remain visually similar to the adopted $N=3$ morphology template.}
\label{fig:order_overlay}
\end{figure}

Figure~\ref{fig:order_overlay} shows the $N=3$--6 templates over the phase-folded observations. The curves are visually very similar over most of the pulsation cycle. The higher-order fits slightly modify the extrema but do not change the main empirical morphology. We therefore adopt $N=3$ for the main benchmark values.

\section{Bootstrap Uncertainty Estimates}

To estimate the statistical stability of the adopted Fourier morphology
parameters, we performed a non-parametric bootstrap analysis of the cleaned
phase-folded Johnson-$V$ data. The cleaned dataset was resampled with
replacement 2000 times. For each bootstrap realization, the adopted
third-order Fourier model was refit and the corresponding amplitudes,
Fourier ratios, Fourier phase combinations, template amplitude, rise
fraction, asymmetry index, and residual statistics were recomputed. The
reported uncertainties correspond to the 16th and 84th percentiles of the
bootstrap distributions.

The bootstrap results are summarized in Table~\ref{tab:bootstrap}. The
template amplitude is stable at the few hundredths of a magnitude level,
with
\[
\Delta V = 0.8544^{+0.0225}_{-0.0222}\ {\rm mag}.
\]
The rise fraction and asymmetry index are also well constrained:
\[
f_{\rm rise}=0.2885^{+0.0080}_{-0.0070},
\]
and
\[
A_{\rm asym}=0.4230^{+0.0140}_{-0.0160}.
\]
The harmonic ratios are
\[
R_{21}=0.4159^{+0.0303}_{-0.0262},
\]
and
\[
R_{31}=0.1547^{+0.0237}_{-0.0227}.
\]
The larger uncertainty in $\phi_{31}$ compared with $\phi_{21}$ reflects
the weaker third harmonic.

\begin{table}[htbp]
\centering
\caption{Bootstrap uncertainty estimates for the adopted $N=3$ empirical Fourier template. Central values are from the adopted fit, and uncertainties are based on the 16th and 84th percentiles of 2000 bootstrap realizations.}
\label{tab:bootstrap}
\begin{tabular}{lc}
\toprule
Quantity & Value \\
\midrule
$A_0$ & $3.9031^{+0.0064}_{-0.0061}$ \\
$A_1$ & $0.3434^{+0.0083}_{-0.0093}$ mag \\
$A_2$ & $0.1428^{+0.0089}_{-0.0084}$ mag \\
$A_3$ & $0.0531^{+0.0080}_{-0.0075}$ mag \\
$R_{21}$ & $0.4159^{+0.0303}_{-0.0262}$ \\
$R_{31}$ & $0.1547^{+0.0237}_{-0.0227}$ \\
$\phi_{21}$ & $249.99^{+4.90}_{-4.62}$ deg \\
$\phi_{31}$ & $142.27^{+11.11}_{-10.68}$ deg \\
$\Delta V$ & $0.8544^{+0.0225}_{-0.0222}$ mag \\
$f_{\rm rise}$ & $0.2885^{+0.0080}_{-0.0070}$ \\
$A_{\rm asym}$ & $0.4230^{+0.0140}_{-0.0160}$ \\
RMS residual & $0.0970$ mag \\
MAD residual & $0.0809$ mag \\
\bottomrule
\end{tabular}
\end{table}

\begin{figure}[htbp]
\centering
\includegraphics[width=0.95\linewidth]{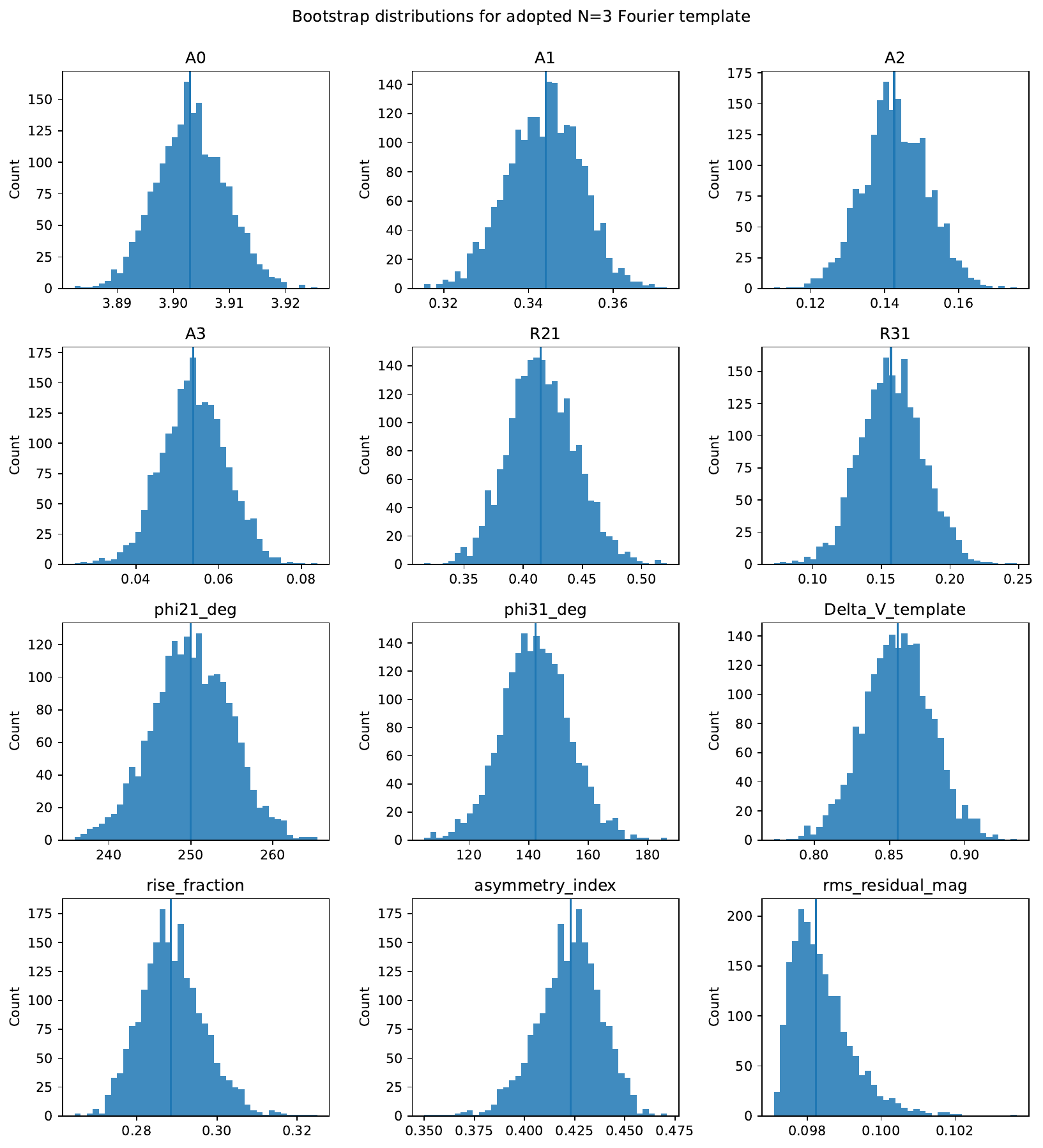}
\caption{Bootstrap distributions for the adopted $N=3$ Fourier template. The distributions show the stability of the main empirical morphology diagnostics under resampling of the cleaned Johnson-$V$ dataset.}
\label{fig:bootstrap}
\end{figure}

\section{Observer-Level Residual Diagnostics}

Because the cleaned AAVSO dataset combines observations from six observers, we examined whether the residuals of the adopted $N=3$ Fourier template show observer-dependent structure. For each observation, we computed
\begin{equation}
\Delta V_{\rm res}=V_{\rm obs}-V_{\rm Fourier}.
\end{equation}
We then grouped the residuals by observer code and computed the mean, median, RMS, and median absolute deviation for each observer.

The results are listed in Table~\ref{tab:observer_residuals}. The four observers with the largest numbers of measurements show systematic mean residuals ranging from approximately $-0.087$ mag to $+0.103$ mag. The difference between the VOL and WGEC mean residuals is nearly $0.19$ mag. This demonstrates that a substantial part of the residual scatter is observer-dependent rather than purely phase-dependent. The smallest observer groups, PGC and HSHC, contain only four and two points, respectively, and therefore should not be overinterpreted. However, they contribute to the largest individual residuals.

\begin{table}[htbp]
\centering
\caption{Observer-level residual diagnostics for the adopted $N=3$ Fourier template. Residuals are defined as $V_{\rm obs}-V_{\rm Fourier}$.}
\label{tab:observer_residuals}
\begin{tabular}{lrrrrr}
\toprule
Observer & $N$ & Mean & Median & RMS & MAD \\
 & & mag & mag & mag & mag \\
\midrule
KPRB & 91 & $+0.0648$ & $+0.0757$ & 0.0817 & 0.0266 \\
VOL  & 76 & $-0.0870$ & $-0.0885$ & 0.0928 & 0.0166 \\
SAH  & 36 & $-0.0331$ & $-0.0272$ & 0.0598 & 0.0338 \\
WGEC & 33 & $+0.1025$ & $+0.1012$ & 0.1054 & 0.0143 \\
PGC  & 4  & $-0.3167$ & $-0.3297$ & 0.3530 & 0.1216 \\
HSHC & 2  & $-0.1032$ & $-0.1032$ & 0.1034 & 0.0064 \\
\bottomrule
\end{tabular}
\end{table}

\begin{figure}[htbp]
\centering
\includegraphics[width=0.90\linewidth]{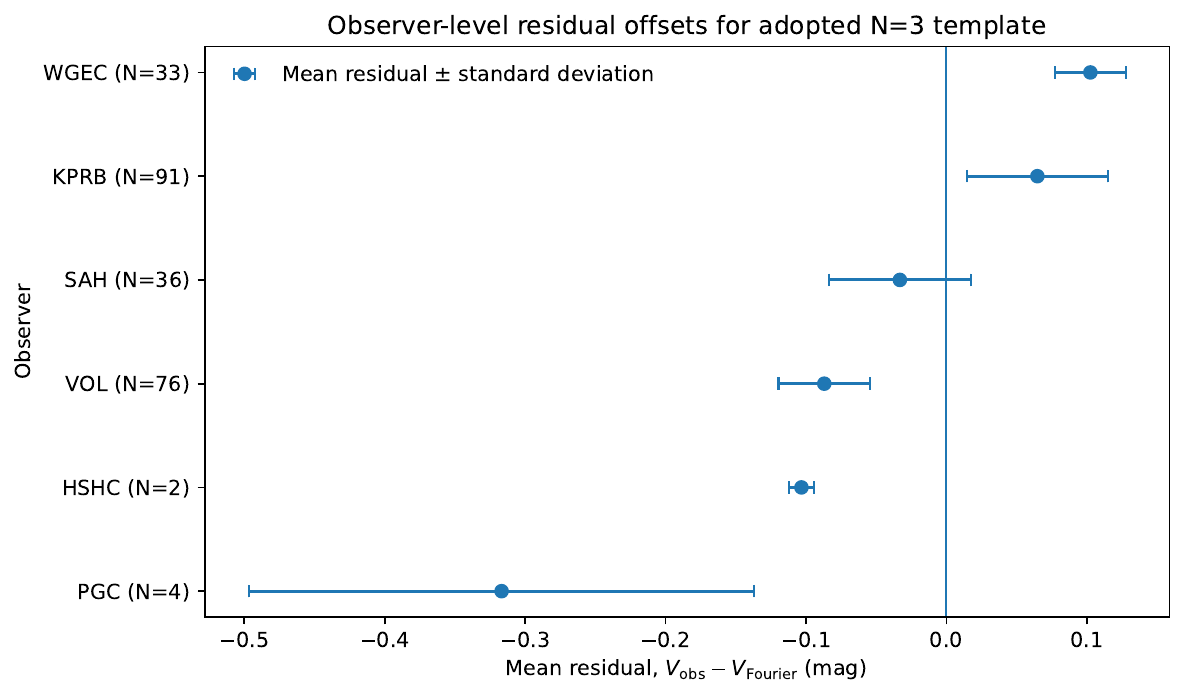}
\caption{Observer-level mean residuals for the adopted $N=3$ Fourier template. Residuals are defined as $V_{\rm obs}-V_{\rm Fourier}$. The figure shows that the residual structure contains observer-dependent offsets at the several-hundredths to $\sim0.1$ mag level for the principal observer groups.}
\label{fig:observer_offsets}
\end{figure}

This observer-level structure supports the interpretation of the adopted template as a low-order empirical morphology benchmark rather than as a precision photometric calibration. The Fourier template describes the mean phase-dependent \Vband\ morphology of the available data, while the residuals retain information about observational heterogeneity, zero-point differences, and individual measurement scatter.

\section{Benchmark Use for Future Model Comparisons}

The empirical template developed here is intended to serve as an observational benchmark for later modeling work. The template can be used to evaluate several different kinds of models:

\begin{enumerate}[leftmargin=*]
    \item nonlinear radial-pulsation models that predict radius, luminosity, effective temperature, and synthetic magnitudes;
    \item atmosphere or bolometric-correction reconstructions that prescribe $R(\phi)$ and $T_{\rm eff}(\phi)$;
    \item semi-empirical SPIPS/Baade--Wesselink reconstructions constrained by photometry, radial velocities, angular diameters, and distance;
    \item stellar oscillation models based on the \textsc{GYRE} code when coupled to an external light-curve or atmosphere calculation;
    \item independent nonlinear hydrodynamic Cepheid calculations.
\end{enumerate}

The relevant comparison quantities are not limited to the period. A useful model comparison should also consider $\Deltav$, $\Rtwone$, $\Rthreeone$, $\phitwone$, $\phithreeone$, $\frise$, $\Aasym$, and the phase dependence of the residuals. In this sense, the present paper defines what the optical light curve looks like before asking which physical model can reproduce it.

This distinction is important. A physical model may reproduce the observed period but still produce a light curve that is too sinusoidal, too low in amplitude, too symmetric, or incorrect in harmonic content. Conversely, an empirical Fourier template may reproduce the observed morphology but does not by itself explain the stellar physics. The two approaches are complementary.

\section{Limitations}

Several limitations should be kept in mind.

First, the dataset is heterogeneous. It contains observations from multiple observers, and the observer-level residual analysis shows offsets at the several-hundredths to approximately $0.1$ mag level for the main observer groups. These offsets are a limitation for precision photometric interpretation, but they also justify the paper's emphasis on a mean empirical morphology template rather than individual-point fitting. Second, the time span is nearly one year, while \dcep\ is known to be a pulsating star with a long observational history. The present work does not attempt a period-change or $O-C$ analysis. Third, the adopted template is a low-order Fourier representation. It is designed to summarize the dominant morphology, not to model every individual point. Fourth, the paper is restricted to \Vband\ photometry. Other bandpasses may have different amplitudes and morphology.

These limitations define the intended use of the result. The template should be used as an empirical \Vband\ morphology benchmark, not as a complete physical model of \dcep.

\section{Conclusions}

We constructed an empirical \Vband\ Fourier template for the observed light curve of \dcep. The main conclusions are:

\begin{enumerate}[leftmargin=*]
    \item The cleaned AAVSO sample contains 242 \Vband\ observations spanning 355.09259 d.
    \item The light curve was folded using $\Pobs=5.366531\,{\rm d}$ and an empirical bright-maximum epoch of JD 2460851.395800.
    \item A third-order Fourier model is adopted as the preferred low-order empirical morphology template.
    \item The adopted template gives $A_0=3.9031^{+0.0064}_{-0.0061}$, $A_1=0.3434^{+0.0083}_{-0.0093}$ mag, $A_2=0.1428^{+0.0089}_{-0.0084}$ mag, and $A_3=0.0531^{+0.0080}_{-0.0075}$ mag.
    \item The corresponding Fourier ratios are $\Rtwone=0.4159^{+0.0303}_{-0.0262}$ and $\Rthreeone=0.1547^{+0.0237}_{-0.0227}$.
    \item The Fourier phase combinations are $\phitwone=4.363117$ rad $(249.99^{+4.90}_{-4.62})^\circ$ and $\phithreeone=2.483084$ rad $(142.27^{+11.11}_{-10.68})^\circ$.
    \item The full template amplitude is $\Deltav=0.8544^{+0.0225}_{-0.0222}$ mag.
    \item The rise fraction is $\frise=0.2885^{+0.0080}_{-0.0070}$, corresponding to an asymmetry index of $\Aasym=0.4230^{+0.0140}_{-0.0160}$.
    \item Fourier orders $N=4$--6 reduce the RMS residual by only approximately 0.0012 mag relative to the $N=3$ model, so the simpler third-order representation is retained.
    \item Observer-level diagnostics show that the residuals contain systematic offsets among the main observer groups, with mean residuals ranging from $-0.0870$ to $+0.1025$ mag. This supports interpreting the template as a mean empirical morphology benchmark rather than a precision photometric calibration.
\end{enumerate}

The final product of this paper is not a physical stellar model. It is a reproducible empirical description of the observed \Vband\ light-curve morphology of \dcep. This template provides a fixed observational target for future comparison with nonlinear pulsation models, synthetic photometry, semi-empirical Baade--Wesselink/SPIPS solutions, \textsc{GYRE}-supported stellar models, and independent hydrodynamic Cepheid calculations.

\section*{Data and Code Availability}

The observational data used in this study were obtained from the AAVSO International Database \citep{aavso_aid}. The analysis workflow consists of scripts for data cleaning, phase folding, Fourier fitting, residual calculation, Fourier-order comparison, bootstrap uncertainty estimation, observer-level residual diagnostics, and figure generation. The final frozen results used in this manuscript are stored in the project directory under \texttt{final\_results\_N3/}.

\section*{Acknowledgments}

This research made use of observations from the AAVSO International Database contributed by observers worldwide \citep{aavso_aid,aavso_data_guidelines}. We acknowledge the AAVSO and its observers for making these data available. We also acknowledge the broader Cepheid literature on Fourier decomposition of observed light curves, which provides the methodological basis for the empirical morphology diagnostics used here.

\bibliographystyle{unsrtnat}
\bibliography{paper7}

\end{document}